\documentclass[%
 reprint,
superscriptaddress,
 amsmath, amssymb,aps,
floatfix,
]{revtex4-1}

\usepackage{graphicx}
\usepackage{dcolumn}
\usepackage{bm}
\usepackage[colorlinks=true,citecolor=blue,linkcolor=magenta]{hyperref}
\usepackage{xr}
\usepackage[utf8]{inputenc}
\usepackage{url}
\usepackage[normalem]{ulem}
\usepackage{upgreek}
\usepackage[mathlines]{lineno}
\usepackage{amsmath}
\usepackage{comment}
\usepackage{microtype}

\begin{document}
\title{MgO-doped lithium niobate nanophotonics for dissipative Kerr soliton microcomb generation}
\author{Tuo Liu}

\author{Biyan Zhan}

\author{Xinyan Chi}

\author{Haoxuan Zhang}

\affiliation{Beijing Engineering Research Center of Mixed Reality and Advanced Display, MIIT Key Laboratory of Photonics Information Technology, School of Optics and Photonics, Beijing Institute of Technology, Beijing 100081, China}
\author{Xianwen Liu}  
\affiliation{Beijing Engineering Research Center of Mixed Reality and Advanced Display, MIIT Key Laboratory of Photonics Information Technology, School of Optics and Photonics, Beijing Institute of Technology, Beijing 100081, China}
\affiliation{corresponding author: xianwen.liu@bit.edu.cn}

\renewcommand{\figurename}{Figure}

\begin{abstract}
\section{Abstract}
Thin-film lithium niobate (TFLN) is a promising platform for integrated photonics, but its relatively  low photorefractive damage threshold limits the device reliability at high pump powers. Magnesium oxide (MgO) doping substantially mitigates this issue, yet the feasibility of MgO-doped TFLN for dissipative Kerr soliton (DKS) microcomb generation has remained unexplored. Here, we demonstrate DKS microcombs in MgO-doped TFLN microring resonators utilizing both air- and silica-cladded geometries. The engineered microresonators exhibit anomalous dispersion and achieve high intrinsic quality factors of up to 6.3\,$\times$\,$10^6$. In contrast to undoped TFLN, we show that the MgO-doped devices exhibit a weak photorefractive response and a pronounced pyroelectric-induced resonance drift. By slowly tuning the pump laser frequency, we generate single-soliton microcombs and perfect soliton crystals in both air- and silica-cladded devices. From the captured DKS spectra, we extract a Kerr nonlinear refractive index of $\sim$2.5\,$\times$\,$10^{-19}$\,m$^2$/W. Our work establishes MgO-doped TFLN as a potential platform that combines high optical damage thresholds with oxide-cladded electro-optic integration capability for advanced microcomb applications.

\end{abstract}

\maketitle
\section{Introduction}
Dissipative Kerr soliton (DKS) microcombs have emerged as a powerful technology for enabling advanced photonic systems \cite{wang2020advances}, with applications ranging from precision metrology~\cite{riemensberger2020massively} and high-capacity telecommunications~\cite{corcoran2025optical} to massively parallel optical computing~\cite{feldmann2021parallel}. These chip-scale frequency combs are generated from continuous-wave-driven, high-quality-factor ($Q$) optical microresonators through the interplay of Kerr nonlinearity and parametric gain~\cite{kippenberg2018dissipative}. Compared to traditional bulk mode-locked lasers~\cite{udem2002optical}, DKS microcombs offer the potential for substantial reductions in size, weight, power, and cost (SWaP-C). The tight spatial confinement and long photon lifetimes afforded by high-Q microresonators dramatically enhance light-matter interactions, reducing the comb generation threshold to the milliwatt level~\cite{zhang2019sub}. This low-power requirement is particularly important for direct integration with electrically pumped diode lasers for turnkey DKS microcomb generation~\cite{shen2020integrated,xiang2021laser,lv2025electrically}.

Among diverse material platforms explored for DKS microcomb generation~\cite{pfeiffer2017octave,gong2020near,liu2021aluminum,wang2021high,cai2025stable,zhu2025octave}, thin-film lithium niobate (TFLN) features commercial availability of high-quality wafers and the potential for monolithic integration of multiple photonic functions \cite{qi2020integrated}. This versatility stems from its combined $\chi^{(2)}$ and $\chi^{(3)}$ optical nonlinearities, pronounced Pockels effect ($\gamma_{33}$\,$\approx$\,30\,pm/V), and amenability to ferroelectric domain engineering~\cite{zhu2021integrated}. Unlike purely Kerr nonlinear platforms, TFLN not only facilitates photorefractive-assisted, self-starting DKS microcombs that bypass complex laser-tuning schemes~\cite{he2019self}, but also unlocks opportunities to integrate electro-optic modulation and $\chi^{(2)}$ frequency conversion within microcomb devices~\cite{wang2019monolithic,gong2022monolithic,he2023high,lu2023two,tang2024}. Typically, DKS microcombs are realized in Z‑cut TFLN microresonators, whose in-plane isotropy enables reliable dispersion engineering of the fundamental TE$_{00}$ modes~\cite{gong2019soliton,song2024octave,zhan2026high}. More recently, DKS operation has also been achieved in X-cut TFLN microresonators by suppressing the strong Raman effect~\cite{song2025stable,nie2025soliton,song2026high}. Despite these advances, prior demonstrations have relied exclusively on air-cladded TFLN microresonator, becasue depositing silicon dioxide (SiO$_2$) can introduce substantial optical losses and exacerbate photorefractive effects~\cite{xu2021mitigating}. This air-cladded requirement complicates on-chip integration of metallic electrodes for active control, especially in Z-cut TFLN microresonators~\cite{he2023high}. Furthermore, congruent TFLN suffers from a relatively low optical damage threshold due to photorefractive-induced index fluctuations and charge-trapping effects \cite{LangePhotorefraction2026}, which compromises the long-term stability and reliability of microcomb devices operating under high intracavity pump powers. 

\begin{figure*}[!ht]
	\centering
	\includegraphics[width=\linewidth]{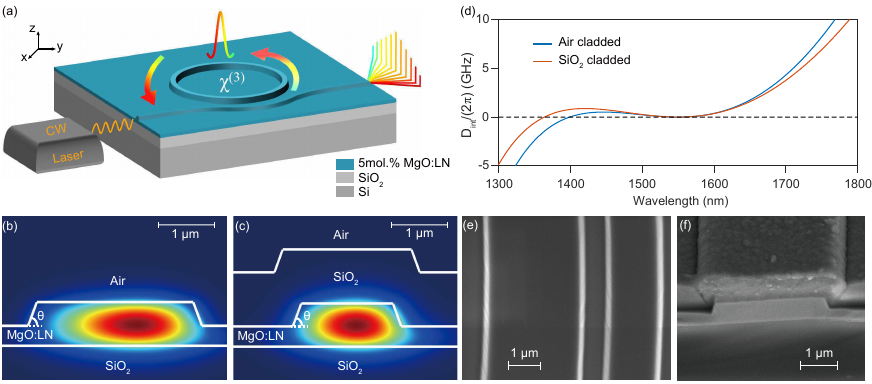}
	\caption{\textbf{Design and fabrication of the MgO:LN microresonators}. (a) Schematic of the device configuration for DKS microcomb generation. (b) and (c) Simulated TE$_{00}$ mode profiles at 1550\,nm for the air- and SiO$_2$-cladded devices, respectively (film thickness: 800 nm; etch depth: 425\,nm; sidewall angle: $\theta$\,=\,70$^\circ$). (d) Calculated integrated dispersion ($D_\mathrm{int}$) curves for the air- and SiO$_2$-cladded devices. (e) and (f) Scanning electron microscopy (SEM) images  showing the waveguide-resonator coupling region of the air-cladded device (e) and the cleaved waveguide facet of the SiO$_2$-cladded device (f).}
	\label{fig1}
\end{figure*}

Magnesium oxide (MgO) doping has been shown to enhance the optical damage threshold of lithium niobate by more than two orders of magnitude~\cite{nakamura2002optical}. Additionally, MgO-doped lithium niobate (MgO:LN hereafter) exhibits a significantly reduced coercive field, which has facilitated periodic poling of nanophotonic chips for high efficiency second-harmonic generation, multi-octave optical parametric oscillation, ultrabroadband supercontinuum generation and self-referencing~\cite{sekine2025multi,wang2018ultrahigh,jankowski2020ultrabroadband,chi2026chip}. Despite these compelling benefits, DKS microcomb generation has remained elusive in MgO:LN microresonators. 

In this work, we demonstrate DKS microcomb generation in Z-cut MgO:LN microring resonators with  either air or  SiO$_2$ top cladding. The engineered resonators exhibit anomalous dispersion in the telecom-C band and achieve a high intrinsic $Q$-factor ($Q_{0}$) of up to 6.3\,$\times$\,10$^6$. Our characterization reveals that both cladding configurations exhibit remarkably weak photorefractive effects, while silica cladding effectively suppresses the pyroelectric-induced resonance wavelength drift observed in air-cladded devices. Akin to self-starting DKS generation in Z-cut TFLN microresonators, we achieved single-soliton microcombs and perfect soliton crystals  in both air-and SiO$_2$-cladded devices by slowly tuning of the pump laser frequency. We further  extract a Kerr nonlinear refractive index of $\sim$2.5\,$\times$\,10$^{-19} \mathrm{m^2/W}$, which notably surpasses the reported value of $\sim$1.7\,$\times$\,10$^{-19} \mathrm{m^2/W}$ for undoped TFLN~\cite{shams2022reduced}. 

\section{Methods}
\noindent \textbf {Device design.} Figure~\ref{fig1}(a) schematically illustrates our device configuration for DKS microcomb generation. The microring resonator is fabricated from a material stack comprising a Z-cut, 5\,mol.\% MgO:LN thin film atop a 2-$\mu$m-thick SiO$_2$ buried layer and a 525-$\mu$m-thick silicon handle substrate. An MgO:LN thickness of 800\,nm is chosen to support anomalous dispersion engineering in both air- and SiO$_2$-cladded structures, while the etching depth is designed to be 425\,nm. As constrained by our negative-tone resist fabrication process~\cite{song2026thin}, the waveguide sidewall angle is set to be $\theta$\,=\,70$^\circ$. The top width for the air- and SiO$_2$-cladded microring resonators is optimized to be 2.8\,$\mu$m and 1.7\,$\mu$m, while their associated central radii are 100\,$\mu$m and 60\,$\mu$m, respectively. For the oxide-cladded configuration, a 1-$\mu$m-thick SiO$_2$ top cladding is adopted. To improve fiber-to-chip coupling efficiency, wide edge couplers are incorporated in both devices by tapering the bus waveguide width to 3\,$\mu$m.

Based on the above resonator parameters, we simulated the intracavity TE$_{00}$ mode profiles using the finite-element method (FEM). As shown in Figs.~\ref{fig1}(b, c), both the designed air- and SiO$_2$-cladded devices achieve strong optical confinement for the TE$_{00}$ mode. The integrated dispersion ($D_\mathrm{int}$) was then calculated by incorporating the wavelength-dependent material dispersion of Z-cut, 5 mol.$\%$  MgO:LN~\cite{zelmon1997infrared}. As shown in  Fig.~\ref{fig1}(d), the designed resonator parameters yield $D_\mathrm{int}$ curves with anomalous dispersion in the telecom-C band, which is essential for the formation of bright DKS microcombs. While the present design only supports phase-matched dispersive waves (DWs) near 1400\,nm, the $D_\mathrm{int}$ profile can be further engineered to facilitate broadband DKS microcomb generation through dual DW emission (see Supplementary Fig.\,S1). When reducing the MgO:LN film thickness to 600\,nm, we note that it is difficult to obtain the desired anomalous dispersion in the  SiO$_2$-cladded MgO:LN microresonator.

\begin{figure*}[!ht]
	\centering
	\includegraphics[width=\linewidth]{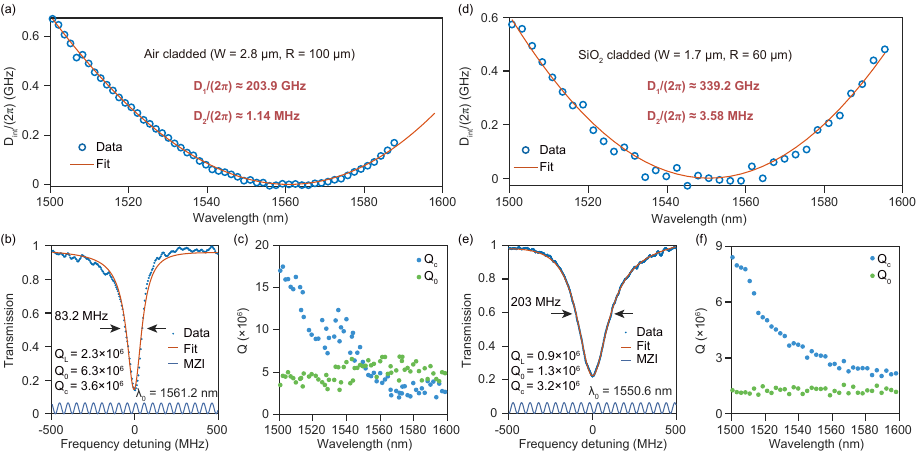}
	\caption{\textbf{$Q$-factor and dispersion characterization}. (a-c) Air-cladded MgO:LN microresonators: (a) Measured $D_{\text{int}}/(2\pi)$ curve with a polynomial fit; (b) normalized resonance spectra near 1561.2\,nm with extracted $Q$-factors; (c) wavelength-dependent $Q$-factors across 1500--1600\,nm. (d-f) SiO$_2$-cladded MgO:LN microresonators with panels corresponding to (a-c) except that the spectra in (e) were recorded near 1550.6\,nm.} 
	\label{fig2}
\end{figure*}

Note that strong Raman-active phonon modes in crystalline platforms such as TFLN can hinder DKS microcomb generation \cite{gong2019soliton}. To suppress stimulated Raman scattering (SRS) in MgO:LN microresonators, we employ dispersive evanescent coupling through concentric pulley couplers, as sketched in Fig.\,\ref{fig1}(a). The pulley geometry controls wavelength-dependent phase matching and modal overlap, enabling selective engineering of external coupling rate \cite{song2024octave}. For both air- and SiO$_2$-cladded devices, we chose a pulley wrap angle of $25^{\circ}$, with optimized bus-waveguide widths (coupling gaps) of 1.4~$\mu$m (1.0~$\mu$m) and 0.8~$\mu$m (1.4~$\mu$m), respectively. As shown in Supplementary Fig.\,S2, these geometries produce wavelength-dependent coupling behavior, with enhanced external coupling toward the Stokes wavelength near 1700\,nm, hence raising the SRS threshold.  

\noindent \textbf {Fabrication.} We prepared two MgO:LN pieces to fabricate the air- and SiO$_2$-cladded devices, respectively. To ensure consistency between the designed and fabricated structures, the MgO:LN film thickness of both pieces was measured to be $\sim$800\,nm by using an optical reflectometer. The optimized microring geometries were patterned by 100-kV electron beam lithography using a negative-tone resist (AR-N 7520.17). The patterns were subsequently transferred into the MgO:LN layer via an angled ion-beam etching (IBE) process~\cite{zhang2026angle}. The waveguide sidewall angle was measured to be $\sim$70$^\circ$ based on our prior study \cite{song2026thin}. The etch depth was characterized to be $\sim$430\,nm, close to the designed value of 425\,nm. After dry etching, the residual resist and redeposited products were removed by wet cleaning. One of the chips was then coated by a 1-$\mu$m-thick SiO$_2$ cladding layer using plasma-enhanced chemical vapor deposition (PECVD). Finally, the two chips were annealed at 500\,$^\circ\text{C}$ for three hours and cleaved to expose the waveguide facets. As shown in Fig.~\ref{fig1}(e), the fabricated resonator exhibits a smooth sidewall profile. The cleaved waveguide facet of the SiO$_2$-cladded device is shown in Fig.~\ref{fig1}(f).

\section{Experimental results}
\subsection {Dispersion and Q-factor characterization.}
We characterized the transmission spectra of the fabricated air- and SiO$_2$-cladded MgO:LN microresonators using the experimental setup detailed in our prior work~\cite{zhang2026angle}. During measurement, a tunable external-cavity laser (Santec TSL-550C) served as the optical source and was swept from 1500\,nm to 1600\,nm at a rate of 100\,nm/s. The off-chip optical power was maintained at a low level of --30\,dBm to suppress unwanted thermal and nonlinear optical effects. The transmitted optical signal was sent to a photodetector (PD) and recorded by an oscilloscope (OSC). To enable reliable extraction of the $D_\mathrm{int}$ profiles and $Q$-factors, we calibrated the frequency axis of the transmission spectra using an unbalanced fiber Mach-Zehnder interferometer (MZI) with a calibrated free spectral range (FSR) of  $\sim$50.9~MHz~\cite{zhang2026all}. As shown in Fig.\,\ref{fig2}(a) and Fig.\,\ref{fig2}(d), polynomial fits to the measured $D_\mathrm{int}$ profiles yielded $D_2/(2\pi) \approx 1.14\,\mathrm{MHz}$ and $D_2/(2\pi) \approx 3.58\,\mathrm{MHz}$ for the air- and SiO$_2$-cladded resonators, respectively, confirming that both devices operate in the anomalous dispersion regime ($D_2>0$). At the pump wavelength, the corresponding FSRs, denoted by $D_1/(2\pi)$, were determined to be 203.9\,GHz and 339.2\,GHz, respectively.

 \begin{figure*}[!ht]
	\centering
	\includegraphics[width=\linewidth]{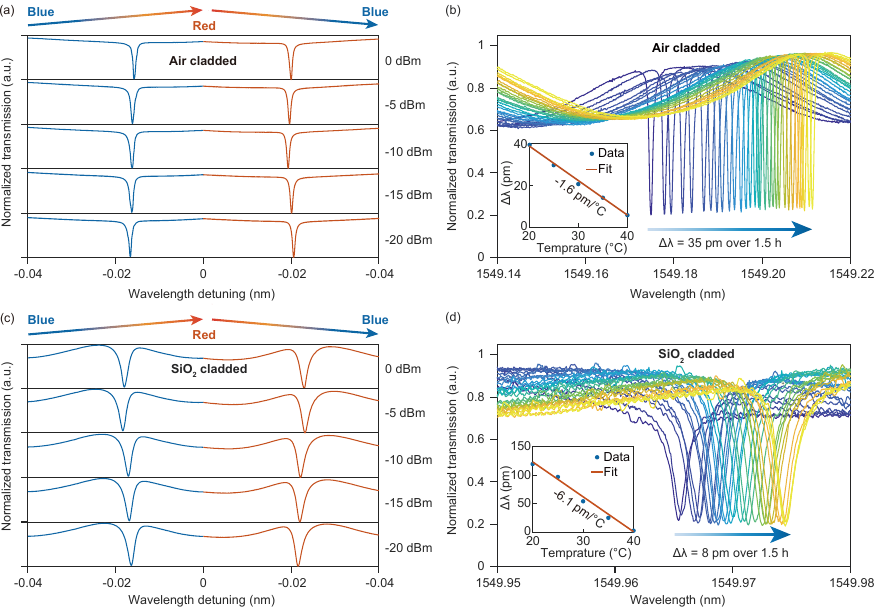}
	\caption{\textbf{Photorefractive and pyroelectric characterization.} (a, b) Air-cladded MgO:LN microresonators: (a) Normalized transmission near 1550 nm measured at on-chip pump powers from $-20$~dBm to $0$~dBm; The blue and orange curves correspond to forward (blue-to-red) and backward (red-to-blue) wavelength scanning, respectively. (b) Characterization of resonance wavelength drift over 1.5 hours at a fixed probe pump power of $-30$~dBm. The chip temperature was raised to and stabilized at 35$^\circ$C. Inset: measured and fitted resonance wavelength shift versus chip temperature at an interval of 3 minutes. Panels (c) and (d) show the corresponding measurements for the SiO$_2$-cladded device.} 
	\label{fig3}
\end{figure*}

Benefiting from a large resonator geometry ($W=2.8\ \mu$m, $R=100\ \mu$m), the air-cladded device achieves a remarkable $Q_0$ of $6.3$\,$\times$\,10$^6$ [Figs.\,\ref{fig2}(b)], corresponding to a low propagation loss of 0.065 dB/cm. This $Q$-factor was extracted at over-coupled conditions, as confirmed by the microresonator's phase-response characterization (see Supplementary Fig.\,S3) . In contrast, the SiO$_2$-cladded device exhibits a relatively lower $Q_0$ value of $1.3$\,$\times$\,10$^6$ [Fig.\,\ref{fig2}(e)]. This reduction is due to the smaller bending radius ($R$\,=\,60\,$\mu$m) and narrower microring width ($W$\,=\,1.7\,$\mu$m) required for anomalous dispersion engineering, which render the resonator more sensitive to sidewall-scattering losses. In addition, hydrogen absorption within the SiO$_2$ cladding deposited via PECVD may also limit the achievable $Q$-factor~\cite{shams2022reduced}. This issue can be mitigated by using a low-temperature, hydrogen-free deposition process \cite{qiu2024Hydrogen}. Nevertheless, the achieved $Q$-factor is sufficient for triggering DKS generation. We further characterized the wavelength-dependent $Q$-factors across the 1500--1600\,nm spectral range.  As shown in Figs.~\ref{fig2}(c) and \ref{fig2}(f), the coupling $Q$-factors ($Q_{\text{c}}$) for both the air- and SiO$_2$-cladded devices decrease with increasing wavelengths, in agreement with our numerical design presented in Supplementary Fig.\,S2. Meanwhile, the extracted $Q_{\text{0}}$ values of both resonators remain relatively constant across the measured spectral range, confirming weak wavelength dependence of the intrinsic optical losses.

\subsection{Characterization of photorefractive and pyroelectric effects}
We evaluated the photorefractive response of the air- and SiO$_2$-cladded microresonators by bidirectionally scanning the laser wavelength across the resonance near 1550\,nm at a rate of 1\,nm/s. The on-chip pump power was varied from --20\,dBm up to 0\,dBm in 5\,dB increments. As depicted in Figs.\,\ref{fig3}(a) and \ref{fig3}(c), both devices exhibit negligible distortion of the resonance lineshapes throughout the investigated optical power range, indicating a weak photorefractive response. This behavior contrasts with the pronounced photorefractive effects previously observed in undoped TFLN microresonators, particularly those with SiO$_2$ claddings~\cite{xu2021mitigating}. 

We further investigated the pyroelectric response of the air- and SiO$_2$-cladded microresonators by tracking their resonance wavelengths under controlled thermal conditions.  The input probe power was maintained at --30 dBm to minimize optical heating and photorefractive effects. The chip temperature was regulated using an external thermoelectric cooler (TEC), while the resonance wavelengths were monitored by repeatedly sweeping a tunable laser across the resonance dip. As shown in Figs.\,\ref{fig3}(b) and \ref{fig3}(d), both devices displayed a gradual resonance redshift after increasing and stabilizing the chip temperature at 35$^\circ\text{C}$. This relaxation stems from the pyroelectric effect inherent to Z-cut TFLN~\cite{ren2025photorefractive}, whereby temperature variations alter the Z-axis spontaneous polarization and generate uncompensated bound polarization charges at the top and bottom waveguide interfaces. The resulting internal electric field modifies the effective refractive index via the Pockels effect, producing a time-dependent resonance shift as charge compensation gradually restores electrostatic equilibrium. Over 1.5 hours, the air-cladded resonator exhibited a substantial redshift of 35 pm, fivefold larger than the 7-pm drift in the SiO$_2$-cladded device. This enhanced stability reflects effective surface passivation by SiO$_2$ cladding, which isolates the waveguide from ambient ionic species and suppresses interfacial charge accumulation and migration~\cite{yin2025fundamental}. 

Using the laser-sweeping approach described above, we characterized the temperature-dependent resonance dynamics by raising the TEC setpoint from 20\,$^\circ\text{C}$ to 40\,$^\circ\text{C}$ in 5\,$^\circ\text{C}$ increments. At each step, approximately 3 minutes were required to stabilize the temperature and readjust the fiber-to-chip coupling. As shown in the insets of Figs.~\ref{fig3}(b) and \ref{fig3}(d), both devices exhibited a net resonance blueshift, yielding effective thermal tuning rates of --1.6\,pm/$^\circ\text{C}$ and --6.1\,pm/$^\circ\text{C}$ for the air- and SiO$_2$-cladded resonators, respectively. These shifts are attributed primarily to the pyroelectric effect, which induces an initial resonance blueshift upon heating~\cite{yin2025fundamental}, followed by a gradual redshift as charge redistribution screens the pyroelectric field. The smaller net blueshift in the air-cladded resonator is consistent with its stronger relaxation that partially offsets the initial blueshift, as supported by its larger long-term redshift of 35\,pm over 1.5\,h [Figs.\,\ref{fig3}(b)]. The thermo-optic contribution is expected to be small because the  intrinsic thermo-optic coefficient relevant to the TE$_{00}$ mode in Z-cut MgO:LN resonators is nearly zero~\cite{moretti2005temperature}.

\begin{figure*}[!ht]
	\centering
	\includegraphics[width=\linewidth]{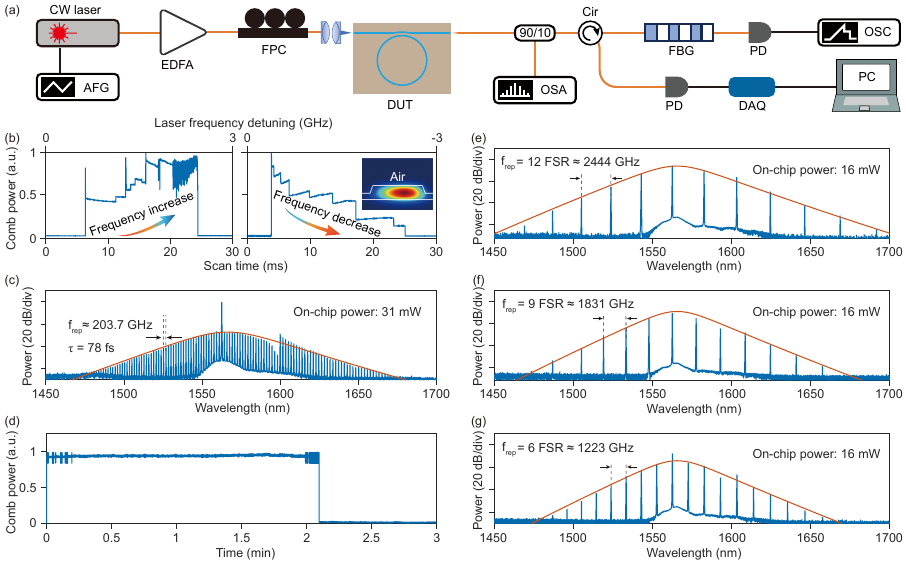}
	\caption{\textbf{DKS generation in the air-cladded MgO:LN microresonator}. (a) Schematic of the experimental setup. Cir: circulator; PD: photodetector; DUT: device under test; DAQ: data acquisition; PC: personal computer. (b) Intracavity comb power evolution during backward (left) and forward (right) laser wavelength sweeps across the resonance near 1561.2\,nm. Inset: simulated TE$_{00}$ cavity mode profile. (c) Single-soliton microcomb spectrum fitted with a $\text{sech}^2$ function (red solid line). (d) Duration of the free-running DKS microcomb. (e--g) Optical spectra of perfect soliton crystals exhibiting extended repetition rates of $f_{\text{rep}}$\,=\,12\,FSR,  9\,FSR, and 6\,FSR, respectively. The red solid lines show corresponding $\text{sech}^2$ envelope fits.}
	\label{fig4}
\end{figure*}

\subsection{DKS microcomb generation in air- and SiO$_2$-cladded microresonators}
Figure~\ref{fig4}(a) presents the experimental setup for DKS microcomb generation~\cite{yin2026wavelength}. A continuous-wave (CW) tunable laser (Santec TSL-550C) served as the pump source, and its output was boosted by an erbium-doped fiber amplifier (EDFA). An arbitrary function generator (AFG) drove the laser to enable bidirectional frequency sweeps across the cavity resonance at a rate of 75 GHz/s. A fiber polarization controller (FPC) adjusted the pump polarization to excite the $\text{TE}_{00}$ waveguide mode, and a lens pair coupled the pump light into the microresonator. The transmitted light was collected by a lensed fiber and analyzed using an optical spectrum analyzer (OSA). Its temporal response was monitored by a real-time oscilloscope after suppressing the residual pump using a fiber Bragg grating (FBG).

For the air-cladded device, we first swept the pump laser wavelength at a rate of 1\,nm/s across the resonance near 1561.2\,nm [Fig.\,\ref{fig2}(b)]. Stair-like features appeared in both the blue- and red-detuned regions [Fig.\,\ref{fig4}(b)], indicating the formation of DKS microcombs. As the pump approached resonance, we manually adjusted its wavelength while applying AFG modulation to access the DKS state. The resulting single-soliton microcomb exhibited a characteristic $\mathrm{sech}^{2}$ spectral envelope [Fig.\,\ref{fig4}(c)], with a fitted pulse width of $\tau$\,=\,78\,fs and a measured comb line spacing of $f_{\mathrm{rep}}\approx203.7$~GHz. The spectral dips near 1600~nm may arise from local dispersion perturbations induced by avoided mode crossings. While DW emission near 1400\,nm is predicted in Fig.\,\ref{fig1}(d), it is not observed experimentally due to the weak evanescent coupling of the pulley coupler in this wavelength regime (see Supplementary Fig.\,S2). 

Using the method described in Ref.\cite{chen2026universal}, we calibrated the input coupling loss to be 4.8\,dB (see Supplementary Fig.\,S4), yielding an estimated on-chip pump power of 31\,mW. The free-running DKS state persisted for about 2 minutes [Fig.\,\ref{fig4}(d)]. We attribute this limited stability to the pronounced pyroelectric response observed in Fig.\,\ref{fig3}(b), which leads to a pronounced resonance redshift as pump-induced heat accumulates. Compared with its undoped Z-cut TFLN counterpart, the Z-cut MgO:LN microresonator provides weaker passive self-stabilization against this drift owing to its suppressed photorefractive response and the negligible thermo-optic coefficient associated with the TE$_{00}$ mode [Fig.\,\ref{fig3}]. Nevertheless, the stability could be improved by using an auxiliary laser to pump an orthogonal transverse magnetic (TM) mode~\cite{xiao2024optimizing}, leveraging its substantially larger thermo-optic coefficient for thermal locking~\cite{moretti2005temperature}. At reduced on-chip pump power of $\sim$16\,mW, we also accessed perfect soliton crystals by manually varying the pump laser detuning relative the cavity resonance [Figs.\,\ref{fig4}(e)--\ref{fig4}(g)] . These states feature extended comb line spacings of 12, 9, and 6\,FSRs, corresponding to repetition rates of 2444\,GHz, 1831\,GHz, and 1223\,GHz, respectively. In all cases, the measured spectra closely match the theoretical sech$^2$ envelopes, suggesting the versatility of the MgO:LN platform for realizing terahertz-repetition-rate DKS microcombs.

\begin{figure*}[!ht]
	\centering
	\includegraphics[width=\linewidth]{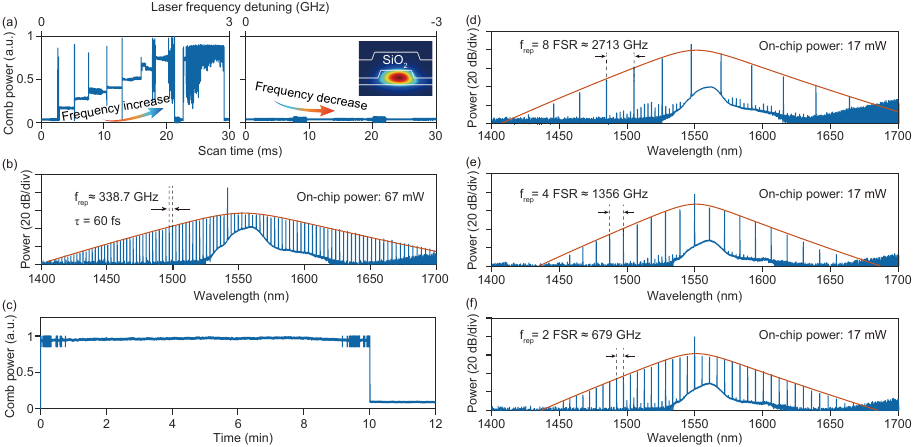}
	\caption{\textbf{DKS generation in the SiO$_2$-cladded microresonator.}	(a) Intracavity comb power evolution during backward (left) and forward (right) laser wavelength sweeps across the resonance near 1550.6\,nm. Inset: simulated TE$_{00}$ cavity mode profile. (b) Single-soliton microcomb spectrum fitted a $\text{sech}^2$ function (red solid line). (d) Duration of the free-running DKS microcomb. (d--f) Optical spectra of perfect soliton crystals exhibiting extended repetition rates of $f_{\text{rep}}$\,=\,8\,FSR,  4\,FSR, and 2\,FSR, respectively. The red solid lines show corresponding $\text{sech}^2$ envelope fits.}
	\label{fig5}
\end{figure*}

We next investigated DKS generation in the SiO$_2$-cladded device with a calibrated input coupling loss of 4.1\,dB (Supplementary Fig.\,S4). As shown in Fig.\,\ref{fig5}(a), at an on-chip pump power of 67\,mW, characteristic stair-like steps appeared in the transmission trace during a backward laser wavelength sweep across the cavity resonance near 1550.6\,nm [Fig.\,\ref{fig2}(f)], indicating the formation of DKS states. In contrast, no discernible DKS steps were observed during a forward laser wavelength sweep across the same resonance. This directional dependence contrasts with the air-cladded device, in which DKS steps were accessible in both tuning directions [Fig.\,\ref{fig4}(b)]. We attribute this difference primarily to the pronounced pyroelectric-induced resonance blueshift in the SiO$_2$-cladded device [inset of Fig.\,\ref{fig3}(d)], which counteracts the forward laser wavelength scan and drives the resonance away from the pump, hence preventing access to the DKS regime. 

Figure\,\ref{fig5}(b) displays the measured single-soliton microcomb spectrum, with an extracted $f_{\mathrm{rep}}$\,$\approx$\,338.7\,GHz and an estimated pulse duration of $\tau$\,=\,60\,fs obtained from a $\mathrm{sech}^{2}$ fit to the spectral envelope. The DKS state persists for more than 10~minutes without active feedback control [Fig.\,\ref{fig5}(c)], substantially longer than that observed in the air-cladded device [Fig.\,\ref{fig4}(d)]. The enhanced DKS stability is consistent with the reduced long-term pyroelectric resonance drift in the SiO$_2$-cladded resonator [Fig.~\ref{fig3}(d)], together with the enhanced thermo-optic response resulting from the higher intracavity thermal load at a moderate $Q$-factor [Fig.\,\ref{fig2}(e)]. At a reduced on-chip pump power of 17\,mW, we observed the formation of perfect soliton crystals by adjusting the pump laser detuning [Figs.~\ref{fig5}(d)--\ref{fig5}(f)], yielding repetition rates corresponding to 8\,FSR ($\sim$2713\,GHz), 4\,FSR ($\sim$1356\,GHz), and 2\,FSR ($\sim$679\,GHz). These states indicates the presence of eight, four, and two equally-spaced solitons circulating in the cavity. For all cases, the recorded optical spectra closely matched the fitted sech$^2$ envelopes.

Based on the DKS microcombs generated in the air- and SiO$_2$-cladded MgO:LN microresonators, we estimated the corresponding Kerr nonlinear refractive index ($n_2$) using the following expression~\cite{yi2015soliton}:
\begin{equation}
n_{2}=\frac{2 c \eta A_{\mathrm{eff}} \left| \beta_2 \right|}{P_{\text{soliton}} Q_{\mathrm{L}}} \frac{1}{\tau}
\label{E1}
\end{equation}
Here, $A_{\text{eff}}$ is the effective mode area,  $c$ is the speed of light in vacuum, $\beta_2$ is the group velocity dispersion, $P_{\text{soliton}}$is the on-chip average DKS power (excluding the pump), and $\tau$ is the DKS pulse width. The waveguide-to-resonator coupling parameter is given by $\eta = Q_{\text{L}}/Q_{\text{c}}$, where  $Q_{\text{L}}$ denotes the loaded $Q$-factor. Notably, $\beta_2$ can be determined from the experimentally extracted dispersion parameters $D_1$ and $D_2$ using:
\begin{equation}
	\beta_{2}=-\frac{n_{g}D_{2}}{cD_{1}^{2}}=-\frac{2\pi D_{2}}{D_{1}^{3}L}
	\label{E2}
\end{equation}
where, $n_g$ and $L$ are the group refractive  index and the physical perimeter of the microresonator, respectively. 

For the air-cladded device, FEM simulations yield $A_{\text{eff}}$\,$\approx$\,1.5\,$\mu\text{m}^2$, while the measured $D_\mathrm{int}$ curve in Fig.\,\ref{fig2}(a) gives $\beta_2$\,=\,--32.17\,ps$^2$/km and the resonance characterization in Fig.\,\ref{fig2}(b) reveals $\eta$\,=\,0.64. Using the calibrated output coupling loss of 6.3\,dB (see Supplementary Fig.\,S4), we extracted $P_{\text{soliton}}$\,=\,0.44\,mW. Substituting these values and the estimated pulse width of $\tau$\,=\,78\,fs [Fig.\,\ref{fig4}(c)] into Eq.\,\ref{E1} yields  $n_2$\,$\approx$\,2.5\,$\times$\,10$^{-19}$\,m$^2$/W. Similarly, for the SiO$_2$-cladded device, FEM simulations yield $A_{\mathrm{eff}}$\,$\approx$\,1.2~$\mu$m$^2$, while Fig.\,\ref{fig2}(d) gives $\beta_2$\,=\,--38.66\,ps$^2$/km and Fig.~\ref{fig2}(e) gives $\eta$\,=\,0.3. Using the calibrated output coupling loss of 4.9\,dB (Supplementary Fig.\,S4), we obtain $P_{\mathrm{soliton}}$\,=\,0.52\,mW. Combining these values with the estimated pulse width of $\tau$\,=\,60\,fs [Fig.\,\ref{fig5}(b)] gives $n_2$\,$\approx$\,2.7\,$\times$\,10$^{-19}$\,m$^2$/W, which is in close agreement with that derived from the air‑cladded device. Our results indicate that MgO doping substantially enhances the Kerr nonlinearity compared to undoped TFLN, which has a reported value of  $n_2$\,$\approx$\,1.7\,$\times$\,10$^{-19}$\,m$^2$/W~\cite{shams2022reduced}. This enhancement is also corroborated by our measurement of an undoped Z-cut TFLN microresonator, for which the same DKS-based extraction method yielded  $n_2$\,$\approx$\,1.6\,$\times$\,10$^{-19}$\,m$^2$/W (see Supplementary Fig.\,S5).

\section{Discussion}
We have demonstrated, for the first time, DKS microcomb generation in MgO:LN microresonators with both air and SiO$_2$ top claddings. The fabricated devices achieved an intrinsic $Q$-factor of up to $Q_0$\,=\,$6.3 \times 10^6$, representing state-of-the-art performance for integrated MgO:LN microresonators. In addition, we observed a strongly suppressed photorefractive response, thereby alleviating the power-handling bottlenecks of undoped TFLN \cite{LangePhotorefraction2026}. Moreover, we show that an SiO$_2$ top cladding effectively suppresses the pyroelectric-induced resonance wavelength drift observed in air-cladded devices. By slowly tuning the pump laser frequency, we accessed single-soliton and perfect soliton-crystal states in both cladding geometries. From the measured DKS characteristics, we extracted a Kerr nonlinear refractive index of $n_2 \approx 2.5 \times 10^{-19}\,\mathrm{m^2/W}$ near 1550\,nm, notably surpassing the value reported in undoped TFLN~\cite{shams2022reduced}. Combining the improved power handling and oxide-cladding compatibility with the intrinsic strong Pockels effect, our results establish MgO:LN-on-insulator as a promising platform for actively controlled, electro-optically reconfigurable DKS microcomb generation.

\def\bibsection{\section{\textbf{references}}}
\bibliographystyle{myaipnum4-1}
\bibliography{ref}

\vspace{1 mm}
\noindent \textbf{Funding.}  National Key Research and Development Program of China (2021YFA1401200); National Natural Science Foundation of China (62575017); Fundamental Research Funds for the Central Universities.

\vspace{1 mm}
\noindent \textbf{Acknowledgment.} 
The authors thank the support from Center of Intelligent Sensing and Precise Testing Technology in Optics and Photonics at Beijing Institute of Technology.

\noindent \textbf{Data availability.} The data that support the findings of this study are available from the corresponding authors upon reasonable request.

\vspace{1 mm}
\noindent \textbf{Competing interests.} The authors declare no competing interests.

 


\end{document}